\documentclass{article}
\usepackage{spconf,amsmath,graphicx}
\usepackage{hyperref}
\usepackage[utf8]{inputenc} %
\usepackage[T1]{fontenc}    %
\usepackage{hyperref}       %
\usepackage{url}            %
\usepackage{booktabs}       %
\usepackage{amsfonts}       %
\usepackage{nicefrac}       %
\usepackage{microtype}      %
\usepackage{graphicx}
\usepackage{makecell}
\usepackage{pifont}%
\usepackage{float}
\restylefloat{table}
\usepackage{amssymb, amsmath}
\usepackage{xcolor}
\usepackage{multirow}
\usepackage{caption}
\usepackage{multibib}
\newcites{supp}{References (Supplementary Material)}
\title{Benchmarking White Blood Cell Classification Under Domain Shift}
\name{Satoshi Tsutsui, Zhengyang Su, Bihan Wen
}
\address{Nanyang Technological University, Singapore}
\begin{document}
\ninept
\maketitle
\begin{abstract}
Recognizing the types of white blood cells (WBCs) in microscopic images of human blood smears is a fundamental task in the fields of pathology and hematology. Although previous studies have made significant contributions to the development of methods and datasets, few papers have investigated benchmarks or baselines that others can easily refer to. For instance, we observed notable variations in the reported accuracies of the same Convolutional Neural Network (CNN) model across different studies, yet no public implementation exists to reproduce these results. In this paper, we establish a benchmark for WBC recognition. Our results indicate that CNN-based models achieve high accuracy when trained and tested under similar imaging conditions. However, their performance drops significantly when tested under different conditions. Moreover, the ResNet classifier, which has been widely employed in previous work, exhibits an unreasonably poor generalization ability under domain shifts due to batch normalization. We investigate this issue and suggest some alternative normalization techniques that can mitigate it. We make fully-reproducible code publicly available\footnote{\url{https://github.com/apple2373/wbc-benchmark}}.
\end{abstract}
\begin{keywords}
White Blood Cells, Leukocytes
\end{keywords}

\section{Introduction}\label{sec:intro}
\begin{figure}[tb]
    \centering
    \includegraphics[width=\columnwidth]{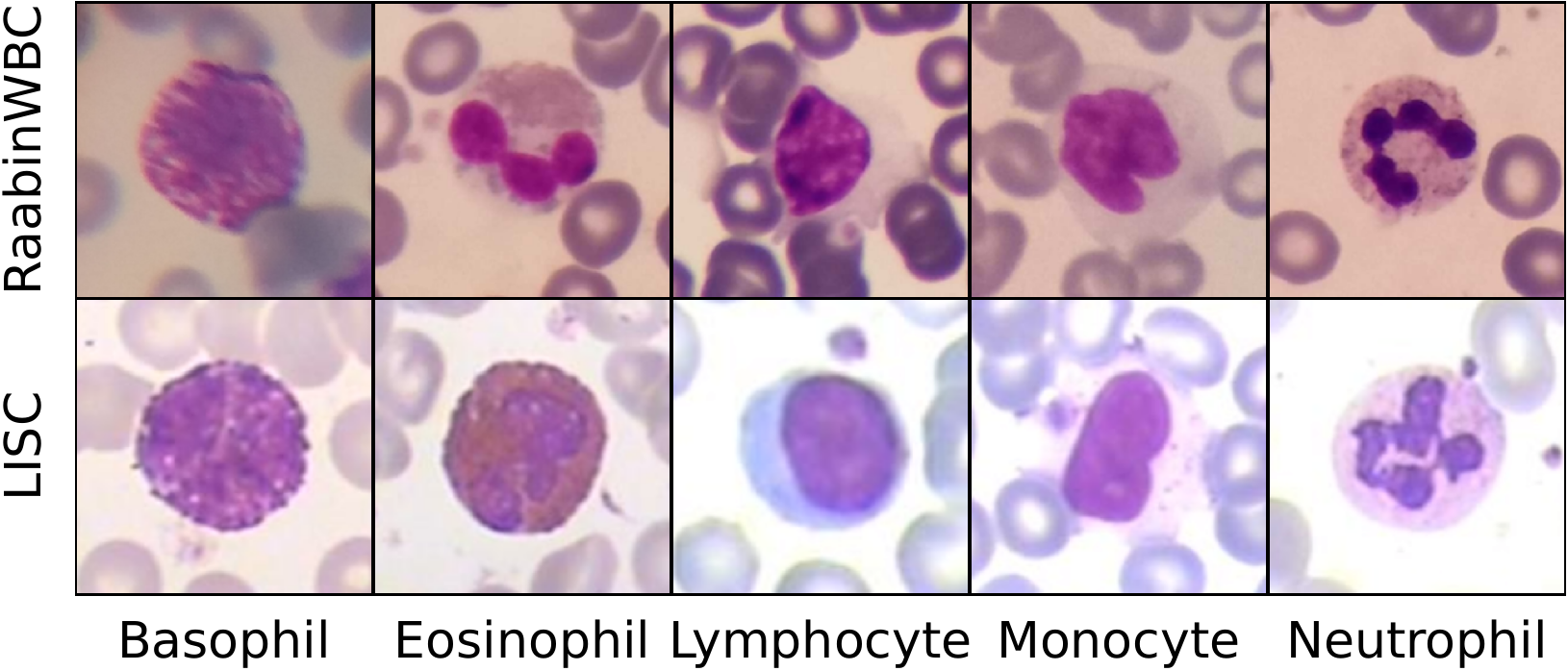}
    \caption{Five types of white blood cells sampled from RaabinWBC~\cite{kouzehkanan2022large}, which we use for training, and LISC~\cite{rezatofighi2011automatic}, which we use for testing domain generalization. See Table~\ref{tbl:dataset} for the dataset statistics.}\label{fig:samples} 
\end{figure}

The microscopic examination of white blood cells (WBCs), also known as leukocytes, in human blood smears is an essential task in the fields of pathology and hematology. This task is particularly important in the diagnosis of blood disorders, such as leukemia, anemia, polycythemia, and immune-related diseases like autoimmune anemia, allergies, and others~\cite{putzu2014leucocyte,kouzehkanan2022large}. Typically, the diagnosis process involves a differential count~\cite{Blumenreich1990TheWB}, which analyzes the distribution of the five types of WBCs, namely neutrophils, eosinophils, basophils, monocytes, and lymphocytes (see Fig.~\ref{fig:samples}). Precise and automated classification of WBCs improves the efficacy of diagnosis.

The recognition of WBCs entails both signal  processing~\cite{bikhet2000segmentation,genovese2021acute,khobragade2015detection,labati2011all,mathur2020mixup} and biomedical expertise~\cite{rezatofighi2011automatic,Mohamed2012AnET,tavakoli2021generalizability,tavakoli2021new,kouzehkanan2022large,jung2022wbc}. A comprehensive literature review on this topic is provided in the survey by Zolfaghari et al.~\cite{zolfaghari2022survey}. Previous studies have made noteworthy contributions to the development of automatic WBC classifiers\cite{almezhghwi2020improved, chen2021transmixnet,saleem2022deep, chen2022accurate} and publicly available datasets~\cite{labati2011all,rezatofighi2011automatic,Mohamed2012AnET,kouzehkanan2022large}. However, until recently, the size of public WBC datasets was limited to only hundreds of images~\cite{labati2011all,rezatofighi2011automatic,Mohamed2012AnET}, which was insufficient to apply state-of-the-art image classification models, such as Convolutional Neural Networks (CNNs). Recently, the RaabinWBC dataset~\cite{kouzehkanan2022large} was released, comprising a relatively large number of WBC images (16k). Subsequent studies~\cite{almezhghwi2020improved, chen2021transmixnet,saleem2022deep, chen2022accurate} have proposed new approaches to improve the classification performance on RaabinWBC.

Meanwhile, the community has paid little attention to benchmarking WBC classification. For instance, the highest accuracy we discovered is 99.91\%~\cite{tavakoli2021generalizability} on RaabinWBC Test-A set using VGG CNN~\cite{vgg} , while another study~\cite{kouzehkanan2022large} reported 98.09\% also using VGG. This difference in performance could be attributed to variations in image preprocessing, data augmentation, or the inherent randomness of stochastic gradient descent. In fact, we found that the range of accuracies can vary by as much as 1\% by simply altering the random seeds in our code (see Fig.\ref{fig:boxplot}), indicating that error bars should be reported. However, we cannot easily compute them for prior work since no paper publishes the implementation for training CNNs on RaabinWBC. Regarding code, some authors have released code for WBC feature extraction \cite{tavakoli2021new} or model weights trained on private datasets \cite{jung2022wbc}, but we are not aware of any work that publishes reproducible code to train baseline CNNs on publicly available WBC classification datasets and report accuracies with error bars.

\begin{table}[tb]
    \caption{Summary of datasets used in our Work. While RaabinWBC~\cite{kouzehkanan2022large} has a large quantity that allows training modern CNNs, we notice that its test set under domain shifts (Test-B) is not suitable for benchmarking domain generalization due to missing WBC types and the heavy imbalance. Therefore, we suggest using another dataset of LISC~\cite{rezatofighi2011automatic} for testing domain generalization. }
    \centering
    \resizebox{\columnwidth}{!}{%
        \begin{tabular}{lccccc|c}
        \toprule
        & Lymph. & Mono. & Neut. & Eos. & Bas. & Total \\
        \midrule
        RaabinWBC Train & 2,427 & 561 &  6,231 & 744 & 212 & 10,175 \\
        \makecell[l]{RaabinWBC Test-A \vspace{-1mm} \\ \small{(Same Microscope)}} & 1,034 & 234 &  2,660 &  322 &  89 &  4,339 \\
        \makecell[l]{RaabinWBC Test-B \vspace{-1mm}\\ \small{(Different Microscope)}} & 148 & 0 & 1971 & 0 &  0	& 2,119 \\
        LISC & 59 & 48 & 56 & 39 & 55 & 257\\
        \bottomrule
        \end{tabular}\label{tbl:dataset}
    }
\end{table}

\textbf{Present Work.} This paper aims to establish solid baselines that can be cited and reproduced by other researchers, which we believe is equally important to developing new methods. Specifically, we demonstrate that standard CNNs can achieve high accuracy (over 98.5\%) without the use of more advanced models (see Sec.\ref{sec:baselines}). However, we have discovered that these high accuracy is only maintained when the model is trained and tested under similar imaging conditions, such as using the same measuring devices or blood processing methods. When the model is tested under domain shifts, which are more realistic scenarios in the real world, the accuracy of ResNet dramatically drops to around 23\% (see Sec.\ref{sec:domainshift}). This finding suggests that the classifier is almost broken, which is surprising given that WBCs only have five categories. Strangely, we find that VGG, which is one generation older than ResNet, can still achieve around 74\% accuracy under domain shifts. This is consistent with previous work that reports similar results~\cite{tavakoli2021generalizability,tavakoli2021new}. Through empirical investigation, we have discovered that batch normalization is the cause of poor generalization and demonstrated that group normalization or pre-trained normalization can mitigate this issue (see Sec.\ref{sec:batchnorm}). Our repaired ResNets perform just as well as VGG and achieve around 74\% accuracy under domain shifts, which to our knowledge, is the highest reported accuracy under the same scenario\cite{tavakoli2021generalizability,tavakoli2021new}. Although 74\% still has room for improvement (see Sec.~\ref{sec:analysis}), we believe that our rigorous benchmark on WBC classification will greatly benefit the community. To facilitate progress, we have open-sourced the implementations$^{1}$ to reproduce our reported results.

\textbf{Contributions.} 1) We establish solid baselines for benchmarking the WBC classification under domain shifts. 2) We empirically demonstrate that group normalization or pre-trained normalization improves the cross-dataset generalization of ResNet, a CNN widely used in previous studies. 3) We open-source the implementation$^1$, which other researchers can fully-reproduce our results on publicly available datasets.

\section{Benchmarking WBC Classification}\label{sec:benchmark}
\subsection{Baselines under Similar Imaging Conditions}\label{sec:baselines}
We aim to establish reproducible WBC classification baselines by training standard image classifiers on publicly available datasets. To accomplish this, we utilize RaabinWBC~\cite{kouzehkanan2022large}, the largest WBC classification dataset, and train well-known CNN models. Our default choice of CNN is ResNet50~\cite{he2016resnet}, which is widely regarded as the most frequently used image classifier and has reportedly received the most citations in the field of artificial intelligence~\cite{crew2019google}. We compute accuracy using the Test-A set, which was obtained from the same microscope and blood processing methods as the Train set. ecause we find that the reported accuracies in the previous literature vary more than 1\% even with the same CNN, we run our implementation 10 times and report 95\% confidence intervals. 

\textit{Implementation Details.} We initialize our CNNs with pretrained weights from ImageNet and optimize them using AdamW~\cite{loshchilov2018decoupled} with a weight decay of 0.005, an initial learning rate of 0.0001, and a cosine learning rate decay~\cite{gotmare2018closer} for 10 (warm-up) + 90 (decay) epochs, totaling 100 epochs. Images are resized to $224 \times 224$ with random horizontal and vertical flips. We intentionally avoid using heavy data augmentation, including color changes, as our goal is to establish baselines rather than maximize performance. Future work is encouraged to explore additional data augmentation on top of our approach. For more details, please refer to our implementation$^1$.

\begin{figure}[tb]
    \centering
    \includegraphics[width=0.7\columnwidth]{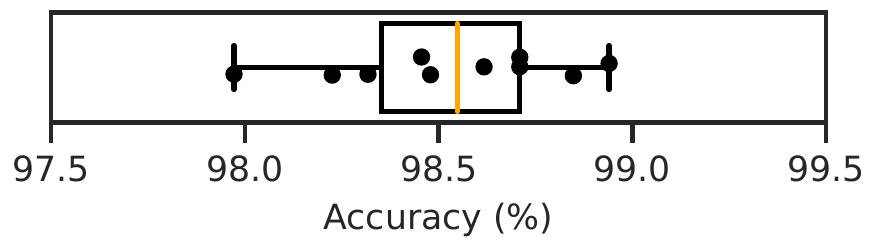}
    \caption{Box plot for ResNet50 classification accuracies (\%) on RaabinWBC Test-A set. To ensure a thorough evaluation, we propose reporting error bars. Although no one has published the code for training CNNs on this dataset, we observed that the reported accuracies in the literature vary by more than 1\% even for the same CNN model. To investigate the source of this variation, we ran our implementation 10 times with different random seeds, where the black dots represent the accuracies of the 10 runs. The mean and 95\% confidence interval is $98.53\pm0.18$. We expect even greater variation among different implementations in previous studies.}\label{fig:boxplot} 
\end{figure}

\textit{Results.} We obtained $98.53\pm0.18$\% accuracy on RaabinWBC Test-A (same microscope) and show the corresponding box plot in Fig.\ref{fig:boxplot}. The highest accuracy is nearly 99\% while the lowest is just under 98\%, spanning intervals of almost 1\%. Given that there are no differences in implementation, we anticipate that the variance of reported results among different papers is likely to be even greater. Since training CNNs inherently involves randomness (due to random weight initialization, order of randomized training samples, etc.), even minor variations in implementation (such as learning rate or preprocessing) can sometimes have a significant impact on relatively small datasets. To address this, we recommend reporting error bars in addition to accuracy. We also experimented with VGG16\cite{vgg}, as it has been previously reported to have the highest accuracy of 99.91\%~\cite{tavakoli2021generalizability}, and obtained an accuracy of $98.75\pm0.06$\%. While this is slightly higher than ResNet's accuracy of $98.53\pm0.18$\%, the confidence intervals overlap.

\definecolor{c0}{rgb}{0.12157, 0.46667, 0.70588}\definecolor{c1}{rgb}{1.00000, 0.49804, 0.05490}\definecolor{c2}{rgb}{0.17255, 0.62745, 0.17255}\definecolor{c3}{rgb}{0.83922, 0.15294, 0.15686}\definecolor{c4}{rgb}{0.58039, 0.40392, 0.74118}
\begin{figure*}[tb]
    \centering
    \includegraphics[height=0.46\columnwidth]{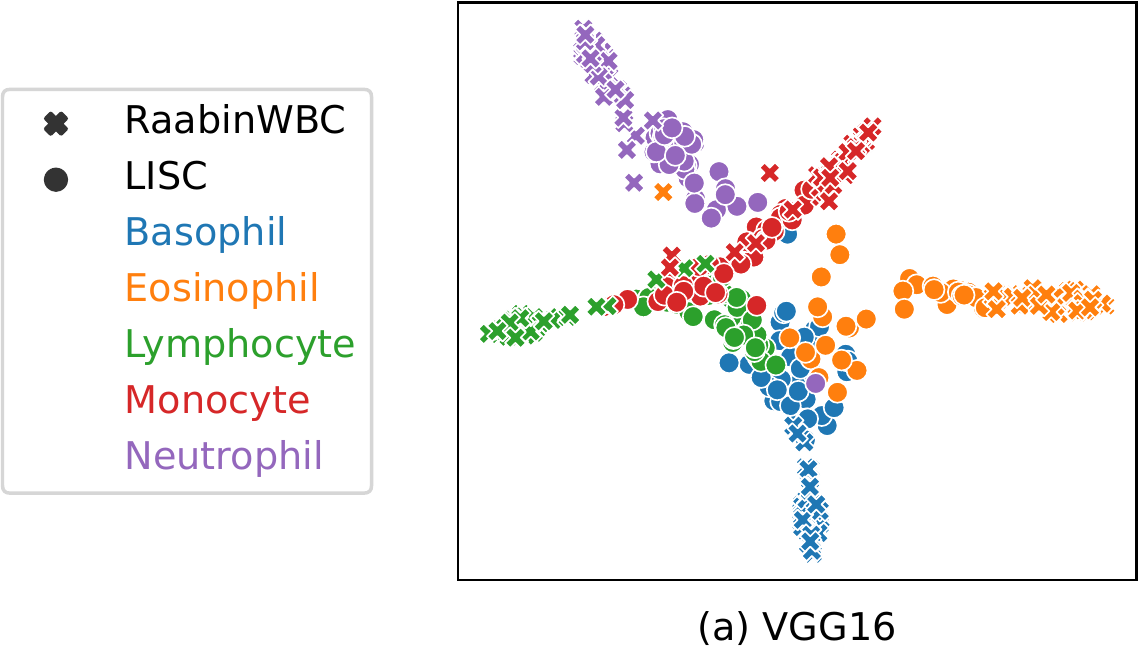}
    \includegraphics[height=0.46\columnwidth]{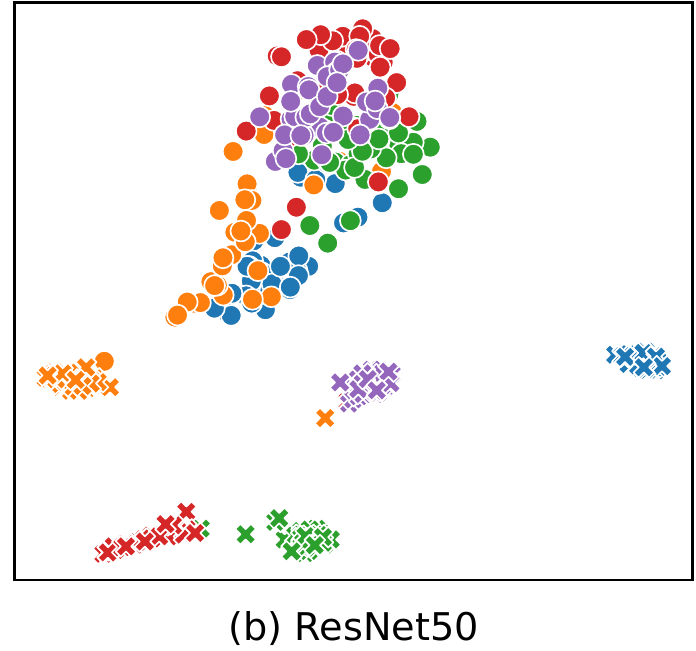}
    \includegraphics[height=0.46\columnwidth]{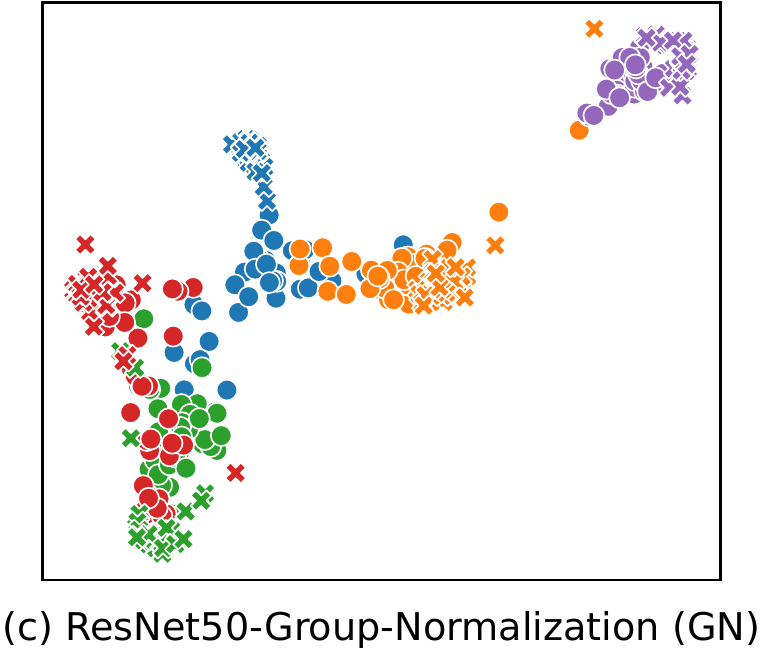}
    \caption{    
     t-SNE~\cite{van2008visualizing} plots show image representations from three CNNs. The markers (\ding{54} and \ding{108}) indicate the datasets while the \textcolor{c0}{c}\textcolor{c1}{o}\textcolor{c2}{l}\textcolor{c3}{o}\textcolor{c4}{r}s represent the WBC types. We sampled 39 images (the smallest category of eosinophils in LISC, as described in Table~\ref{tbl:dataset}) per class and per dataset, and plotted 390 ($= 39\times5\times2$) data points in a 2D space using CNN features trained from RaabinWBC. \textit{(b)}: ResNet50 separates images from RaabinWBC (\ding{54}) and LISC (\ding{108}) more clearly and places them farther apart, even if they belong to the same WBC type (the same color), indicating poor domain generalization ability. \textit{(a, c)}: VGG16 and ResNet50-GN still keep images of RaabinWBC (\ding{54}) and LISC (\ding{108}) relatively close if they belong to the same WBC type (the same color), indicating greater domain generalization ability than ResNet50.
    }\label{fig:tsne} %
\end{figure*}

\subsection{Evaluate under Domain Shift}\label{sec:domainshift}
Since we achieved very high accuracy ($>$98.5\%) when training and evaluating under similar imaging conditions (Test-A), we believe it is appropriate to move on to a more practical scenario where models are trained and evaluated under different imaging conditions. To do so, we examined the RaabinWBC Test-B set, which was collected using a different microscope. However, we realized that it may not be the best dataset for evaluating classifiers under domain shifts due to its composition: it only includes neutrophils and lymphocytes, with neutrophils comprising 93\% (See Table~\ref{tbl:dataset}). As a result, we believe that the LISC dataset~\cite{rezatofighi2011automatic} may be more suitable than the Test-B set, as it is more class-balanced, enabling us to use plain accuracy to compare different methods. One drawback of using the LISC dataset is that it contains much fewer images, but reporting error bars with multiple runs can help mitigate this issue.

\newcommand{\wbccolors}{\definecolor{c0}{rgb}{0.12157, 0.46667, 0.70588}\definecolor{c1}{rgb}{1.00000, 0.49804, 0.05490}\definecolor{c2}{rgb}{0.17255, 0.62745, 0.17255}\definecolor{c3}{rgb}{0.83922, 0.15294, 0.15686}\definecolor{c4}{rgb}{0.58039, 0.40392, 0.74118}\textcolor{c0}{c}\textcolor{c1}{o}\textcolor{c2}{l}\textcolor{c3}{o}\textcolor{c4}{r}}

We propose training models on RaabinWBC and evaluating them with LISC as a practical benchmark for WBC classification. The statistics for the datasets are presented in Table~\ref{tbl:dataset}, and sample images from LISC are displayed in the second row of Fig.\ref{fig:samples}. In comparison to RaabinWBC images (the first row of Fig.\ref{fig:samples}), LISC images have different coloring due to differences in blood processing conditions, such as staining, in addition to different imaging conditions, such as microscopes or cameras. To confirm the existence of domain differences, we use t-SNE~\cite{van2008visualizing} to visualize images from RaabinWBC and LISC using ResNet and VGG, both trained in Sec.\ref{sec:baselines}. The resulting 2D plots in Fig.\ref{fig:tsne}-ab show that the data points for RaabinWBC (\ding{54}) and LISC (\ding{108}) tend to form different clusters, regardless of WBC types (\wbccolors), indicating a domain discrepancy between the two datasets.

\begin{figure}[tb]
    \centering
    \includegraphics[width=\columnwidth]{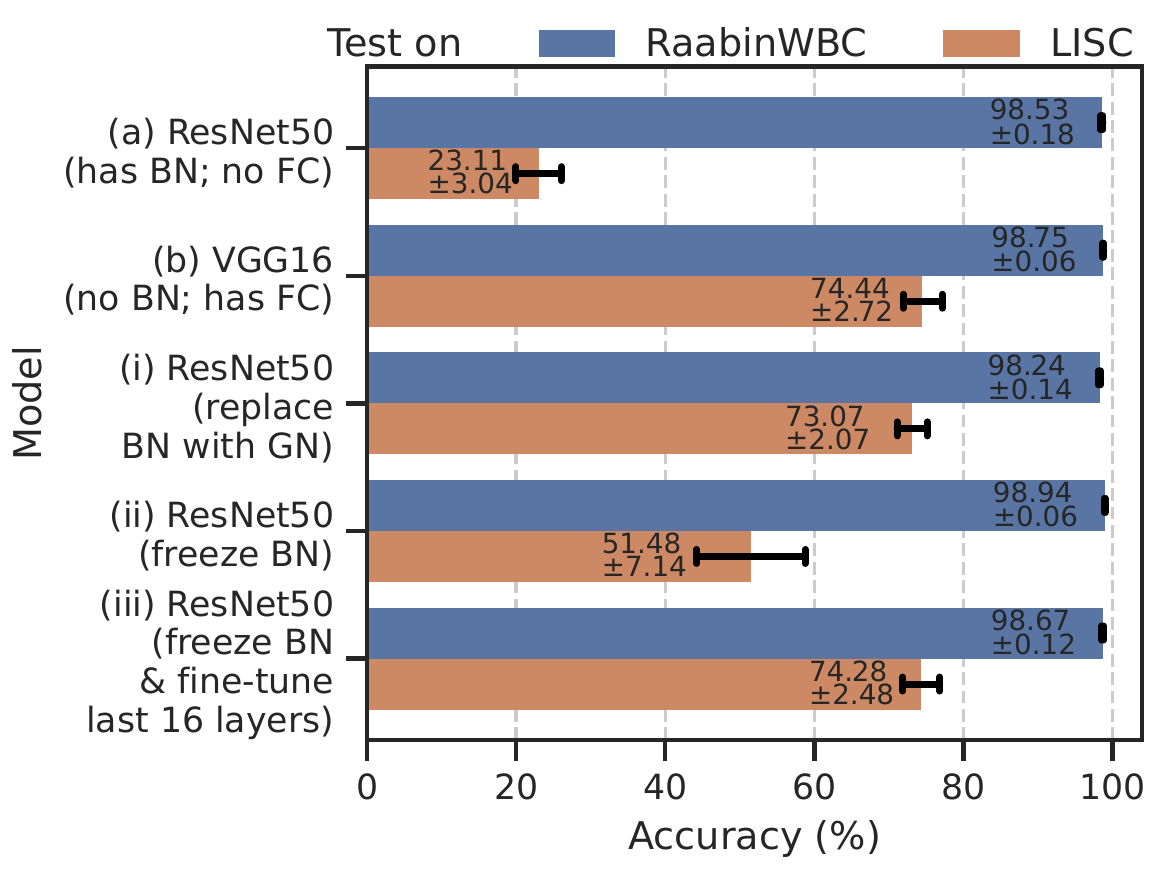}
    \caption{Accuracies of different models trained on RaabinWBC, while tested on RaabinWBC (no domain shift) and LISC (domain shift) datasets. \textit{(a)}: ResNet50 has unreasonable accuracy drop when tested under domain shifts, but \textit{(b)}: VGG does not have it. \textit{(i)}: Replacing Batch Normalization (BN) with Group Normalization (GN) prevents the unreasonable accuracy drop. \textit{(ii)} and \textit{(iii)}: Freezing BN layers and some more layers can also alleviate it. See Sec.~\ref{sec:batchnorm} for details.} \label{fig:barchart} %
\end{figure}

\textit{Results.} We evaluated the two baseline models trained on RaabinWBC with LISC and report their mean accuracies and confidence intervals in Fig.\ref{fig:barchart}-a and -b. The accuracies (\%) for ResNet50 and VGG16 are $23.11\pm3.04$ and $74.44\pm2.72$, respectively. As this task only has five classes, where random guessing can achieve 20\%, ResNet's performance indicates that the classifier is essentially broken under domain shifts. Interestingly, VGG still has a relatively high accuracy, although it still has significant room for improvement compared to the accuracy evaluated under a similar domain. This interesting phenomenon is also supported by the t-SNE plots in Fig.\ref{fig:tsne}-a and -b. In Fig.\ref{fig:tsne}-a, VGG keeps \ding{54} and \ding{108} (i.e., images from different datasets) of the same color (i.e., the same WBC type) relatively close. However, in Fig.\ref{fig:tsne}-b, ResNet separates \ding{54} and \ding{108} more clearly, placing them far apart even if they have the same color, which indicates that the image representations are dramatically changed even if they belong to the same WBC type.

The difference in domain generalization ability between ResNet and VGG is counter-intuitive because ResNet is one generation ahead of VGG and should not be significantly worse than VGG. We infer that an architectural difference between these two models is causing this gap, which we investigate in Sec.~\ref{sec:batchnorm}.

\begin{table}[t]
    \caption{Ablation studies to identify the cause of ResNet's dramatic performance drop on LISC. The results suggest that Batch Normalization (BN) layers adversely affect the domain generalization ability, while Fully-Connected (FC) layers do not. }
    \centering
    \resizebox{\columnwidth}{!}{%
        \begin{tabular}{lccccc}
        \toprule
         &  & \multicolumn{2}{c}{Layer} & \multicolumn{2}{c}{Test Accuracy (\%)} \\
         & Model & FC & BN & RaabinWBC~\cite{kouzehkanan2022large}-A & LISC~\cite{rezatofighi2011automatic} \\
         \toprule
        (a) & \multirow{2}{*}{ResNet} & - & \checkmark & $98.53\pm0.18$  & $23.11\pm3.04$ \\
        (a') &  & \checkmark & \checkmark & $98.45\pm0.19$ & $27.74\pm3.44$ \\
        \midrule
        (b) & \multirow{2}{*}{VGG} & \checkmark & - & $98.75\pm0.06$ & $74.44\pm2.72$ \\
        (b') &  & \checkmark & \checkmark & $98.68\pm0.23$ & $33.74\pm8.04$\\
        \bottomrule
        \end{tabular}\label{tbl:ablation}
    }%
\end{table}

\section{Batch Norm Issue Under Domain Shift}\label{sec:batchnorm}
ResNet's performance is drastically reduced under domain shifts when compared to VGG, as seen in Fig.~\ref{fig:barchart}-a and -b. In this section, we investigate the cause of this issue and explore possible ways to address it.

\textbf{Identifying the cause of the issue.} To identify the building block responsible for the difference in domain generalization between ResNet and VGG, we examine the architectural differences between the models and conduct ablative experiments. We observe that, in addition to the residual connections, \textbf{(a)} \textit{ResNet removed Fully-Connected (FC) layers from VGG}, and that \textbf{(b)} \textit{ResNet added Batch Normalization (BN) layers, which were not present in VGG}. Therefore, we perform two ablation studies on ResNet: \textbf{(a')} \textit{adding FC layers to ResNet}, and \textbf{(b')} \textit{adding BN layers to VGG}. We summarize our observations and ablations in Table~\ref{tbl:ablation}. We note that (b') is a compromised solution as we cannot stably train very deep CNNs without a normalization technique like BN~\cite{bn,gn}. We also note that adding BN to VGG requires ImageNet pretrained weights of VGG with BN, which are available in PyTorch. We do not conduct ablative experiments on the residual connections as they are the essence of ResNet: Res(idual)-Net(work). The results are shown in Table~\ref{tbl:ablation}. The addition of VGG-style fully-connected layers to ResNet, (a) $\rightarrow$ (a'), improves the accuracy (\%) on LISC set from $23.11\pm3.04$ to $27.74\pm3.44$, but is still far behind VGG's $74.44\pm2.72$. On the other hand, the addition of ResNet-style BN layers to VGG, (b) $\rightarrow$ (b'), causes a sudden drop in accuracy on LISC from $74.44\pm2.72$ to $33.74\pm8.04$. This indicates that \textit{batch normalization is the cause of the poor generalization}.

\textbf{Literature about BN's problems.} Now we know that batch normalization (BN)~\cite{bn} is the problem, so we briefly review the related literature to find ways to address it. The BN normalizes the intermediate feature maps of CNNs to have a mean of zero and a standard deviation (std) of one over the training set.  During training, it estimates the mean and std using moving averages from mini-batches of SGD, and during testing, it normalizes the feature maps using the mean and std estimated in training. Intuitively, this makes it harder to generalize to different domains during testing, which has also been reported in a preprint\cite{qing2020invariant} on natural image classification. However, its proposed solution~\cite{qing2020invariant} is unfortunately not applicable to our work, as it trains domain-invariant batch normalization by having multiple domains during training, which we do not have. Additionally, we found a preprint~\cite{bnissue} highlighting BN instability issues due to reliance on the batch for mean/std estimation. A potential solution is Group Normalization (GN),  which normalizes feature maps within images using channel/height/width dimensions instead of batch dimension.

\textbf{Fixing BN's issue.} Inspired by the literature mentioned above regarding BN, we discuss two potential solutions for our generalization issue. Simply removing batch normalization from ResNet is not a viable option, as training a deep network without a technique like BN can be very unstable~\cite{bn, gn}. Additionally, we require ImageNet pretrained weights, which are unavailable for ResNet without BN.\textbf{(i)} The first option is to \textit{Replace BN with GN.}  The idea behind this approach is that if BN normalizes approximately over the training set, making the model dependent on the training domain, then normalizing within each image using GN can make the model independent of the training domain. A disadvantage is the need to prepare ImageNet pretrained weights for ResNet with GN, which fortunately can be avoided since the authors of GN have already published the ImageNet pretrained weights.

\textbf{(ii)} The second option is to \textit{freeze the BN layers with ImageNet pretrained parameters.} The concept behind this approach is that we can prevent overfitting to the training domain by enforcing the use of mean and std from a completely different domain, namely, ImageNet.

\textbf{Repaired ResNet results:} We show the performance of our proposed fixes in Fig.~\ref{fig:barchart}-i and -ii. (a) $\rightarrow$ (i): Replacing BN with GN significantly improved ResNet accuracy (\%) on the LISC dataset from $23.11\pm3.04$ to $73.07\pm2.07$, which is comparable to VGG's $74.44\pm2.72$. (a) $\rightarrow$ (ii): Freezing the BN layers with ImageNet pretrained parameters improved ResNet's accuracy on the LISC set to $51.48\pm7.14$, although still trailing behind VGG's. To address this, we observed that ResNet is deeper than VGG and may be more susceptible to overfitting to a specific domain. Consequently, we experimented \textbf{(iii)} \textit{freezing the first layers and finetuning only the last 16 layers}, the same number as VGG16. This improved ResNet's accuracy on the LISC set to $74.28\pm2.48$, matching VGG's. However, we note that this solution is not based on a solid foundation and is more of a hack, since we cannot explain why ResNet with GN did not require freezing some layers if the deeper layers were the issue.

\section{Performance Analysis}\label{sec:analysis}
\begin{figure}[t]
    \centering
    \includegraphics[width=0.75\columnwidth]{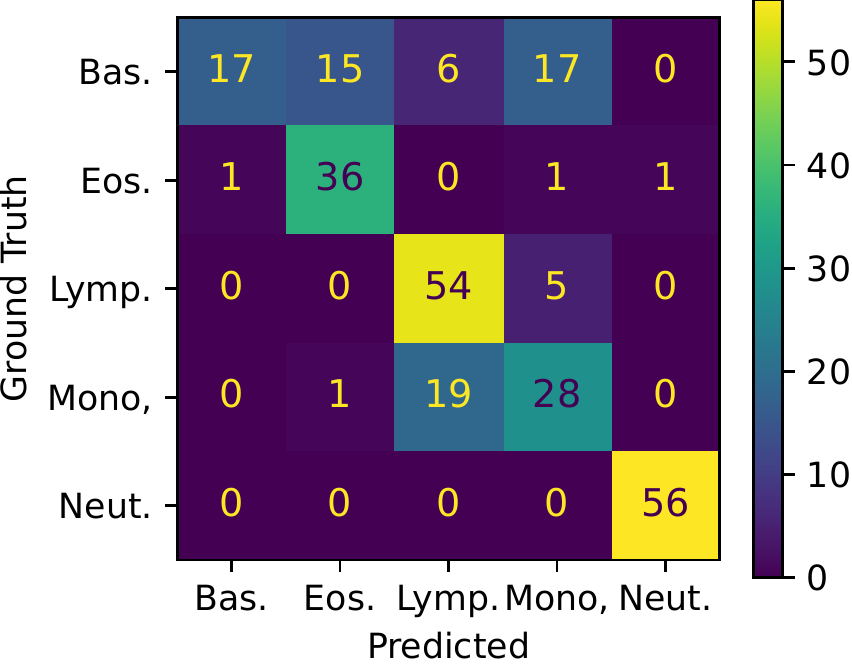}
    \caption{Confusion matrix tested on LISC while trained on RaabinWBC. Basophils (Bas.) is the most challenging WBC to recognize, which may be due to the extremely low quantity (see the first row of Table~\ref{tbl:dataset}) in the training data than other WBC types. }\label{fig:cm} %
\end{figure}

This section examines the performance of the baseline CNN, ResNet50 with group normalization. Out of the 10 trained models, we selected the one with an accuracy of 74.31\% (the closest to the mean) for further analysis and present the precision, recall, and F-measure (harmonic mean of them) in Table~\ref{tbl:results} and the confusion matrix in Fig.~\ref{fig:cm}.

The classifier achieves a high F-measure for neutrophils, and relatively high F-measures for eosinophils and lymphocytes, both of which exhibit lower precision than recall. Monocytes have the second lowest F-measure, with both precision and recall equally low. Monocytes and lymphocytes are frequently confused with each other. Basophils have the worst F-measure, where recall is particularly low and precision is high. Basophils are the most difficult to classify and are often misclassified as eosinophils or monocytes. This poor performance may be due to the heavy class imbalance in the training data (see Table~\ref{tbl:dataset}), reflecting the fact that basophils are the rarest white blood cells, composing only 1\% or less. We observe that the majority of classes tend to have higher F-measures, while the minority classes have lower F-measures, indicating that addressing the heavy class imbalance in the training phase could be future work.

\section{Conclusion}\label{sec:conclusion}
In this work, we established a benchmark for white blood cell (WBC) classification. Building on excellent prior work and publicly available datasets, we demonstrated that standard CNNs perform well when evaluated under imaging conditions similar to the training data. Consequently, we suggest shifting focus towards evaluation under unseen imaging conditions, which is more realistic. We trained baseline models, evaluated them under domain shifts, and identified opportunities for further improvement. We also discovered an issue with batch normalization, a commonly used technique in many baseline CNNs, and proposed ways to address it. We have made our code$^1$ publicly available for full reproducibility of our results.

\vspace{1em}
\textbf{Acknowledgements.} This work was supported in part by Sysmex Corporation, Japan, as well as the MOE AcRF Tier 1 (RG61/22) and Start-Up Grant. It was carried out at the Rapid-Rich Object Search (ROSE) Lab at Nanyang Technological University, Singapore. We thank Winnie Pang for helpful discussions. \nocite{acevedo2020pbc,matek2019human,liu2022convnet,dosovitskiy2021an}

\vspace{1em}
\textbf{Appendix.} We encourage readers to refer to our arXiv preprint version (arXiv:2303.01777), which includes supplementary material.

\begin{table}[t]
    \caption{Precision and Recall on LISC}
    \centering
    \resizebox{0.8\columnwidth}{!}{%
        \begin{tabular}{lrrrrr}
        \toprule
          &   Bas. &   Eos. &  Lymp. &  Mono. &   Neut. \\
        \midrule
        Precision &  94.44 &  69.23 &  68.35 &  54.90 &   98.25 \\
        Recall    &  30.91 &  92.31 &  91.53 &  58.33 &  100.00 \\
        F-measure &  46.58 &  79.12 &  78.26 &  56.57 &   99.12 \\
        \bottomrule
        \end{tabular}\label{tbl:results}
    }%
\end{table}

\clearpage
\vfill\pagebreak

\clearpage
\begin{table}[t!]
    \centering
    \resizebox{\textwidth}{!}{
        \begin{tabular}{clcc}
        \toprule
        {} &                         {} & \multicolumn{2}{c}{Test Accuracy (\%)} \\
        {Index} &                         {Model} & RaabinWBC-A &           LISC \\     
        \midrule
        (a) &                      Default ResNet50 (has batch norm; no fully-connected layers) &   $98.53\pm0.18$ &  $23.11\pm3.04$ \\
        (a') &                   ResNet50 (has batch norm; add fully-connected layers) &   $98.46\pm0.19$ &  $27.74\pm3.44$ \\
        (i) &                   ResNet50 (replace batch norm with group norm; no fully-connected layers) &   $98.24\pm0.14$ &  $73.07\pm2.07$ \\
        (ii)&             ResNet50 (freeze batch norm; no fully-connected layers) &   $98.94\pm0.06$ &  $51.48\pm7.14$ \\
        (iii) &  ResNet50 (freeze batch norm; no fully-connected layers; fine-tune last 16 layers only)  &   $98.67\pm0.12$ &  $74.24\pm2.46$ \\
        (b) &                         Default VGG16 (no batch norm; has fully-connected layers)  &   $98.75\pm0.06$ &  $74.44\pm2.72$ \\
        (b') &                      VGG16 (add batch norm; has fully-connected layers) &   $98.64\pm0.18$ &  $32.33\pm6.17$ \\
        \midrule
        (c) &                           ViT-Base-16 (use layer norm instead of batch norm) &   $98.33\pm0.14$ &  $69.77\pm3.09$ \\
        (d) &                           ConvNeXt-Tiny (similar \# params with ResNet50 but many incremental updates; use layer norm instead of batch norm) &   $98.83\pm0.09$ &  $67.35\pm2.51$ \\
        \bottomrule
        \end{tabular}
    }
    \captionsetup{width=\textwidth}
    \twocolumn[        \caption{Results of ViT and ConvNeXt models along with those presented in the main paper. The Index is the same as in Table~\ref{tbl:ablation} and Fig.~\ref{fig:barchart}.}        \label{tbl:additional}    \vspace{-0.8cm}    ]
\end{table}

\section{Supplementary Materiel}
\subsection{PBC Dataset}
We regret that we did not include the PBC dataset~\citesupp{acevedo2020pbc} in our study. This dataset comprises a total of 17,092 images that can be used for training and testing, while the specific data split employed by the authors is not available. In addition, it includes more classes than the five classes that we used in our experiments. It is worth noting that this dataset is as large as the RaabinWBC dataset~\cite{kouzehkanan2022large}, and was actually published before the RaabinWBC dataset. However, neither dataset cited the other. We became aware of this after the acceptance of our paper, and have included this information here for the sake of completeness.

\subsection{Nature-MI Dataset} 
After finalizing our ICASSP paper, we came across another paper with over 10k WBC images~\cite{matek2019human} published in Nature Machine Intelligence (Nature-MI). This dataset was published even before the RaabinWBC and PBC datasets, but neither of them cited it. This could be due to the paper's focus on a specific type of leukaemia, not on recognizing WBCs in general. Still, the dataset includes 18,365 images, covering the five types of WBCs we use. Importantly, this dataset shows a clear imbalance, for instance, it only contains 79 basophils compared to 8,484 neutrophils. This points to the need for methods that can handle such an imbalance, suggesting an interesting direction for future work.

\subsection{Recent Models without Batch Norm}
In our main paper, we demonstrated that batch normalization makes it difficult for models to generalize beyond the training dataset. To address this issue, we proposed the use of group normalization as an alternative normalization technique that is not dependent on batch. We note that some recent image classification models, including ConvNeXt\citesupp{liu2022convnet} and Vision Transformers~\citesupp{dosovitskiy2021an}, have already abandoned batch normalization in favor of layer normalization, which is a special case of group normalization. We evaluated the performance of these models pretrained on ImageNet and report the results in lines (c) and (d) of Table~\ref{tbl:additional}. As expected, these models did not exhibit a significant drop in accuracy on the LISC dataset, in contrast to the default ResNet model with batch normalization.
\clearpage
\bibliographystyle{IEEEbib}
{\footnotesize
\bibliography{refs}}

\end{document}